\documentclass[conference]{IEEEtran}
\IEEEoverridecommandlockouts
\usepackage{cite}
\usepackage{amsmath,amssymb,amsfonts}
\usepackage{algorithmic}
\usepackage{graphicx}
\usepackage{textcomp}
\usepackage{xcolor}
\usepackage{cite}
\usepackage{bm}
\usepackage{amsfonts}
\usepackage{amssymb}
\usepackage{amsmath}
\usepackage{algorithm}
\usepackage{algorithmic}
\usepackage{array}
\usepackage{tabularx}
\usepackage{subfigure}
\DeclareRobustCommand{\IEEEauthorrefmark}[1]{\smash{\textsuperscript{\footnotesize #1}}}

\def\BibTeX{{\rm B\kern-.05em{\sc i\kern-.025em b}\kern-.08em
    T\kern-.1667em\lower.7ex\hbox{E}\kern-.125emX}}
\begin{document}

\title{5G Direct Position Estimation for Precise Localization in Dense Urban Area\\}

\author{
    \IEEEauthorblockN{Sijia Li, Sergio Vicenzo, and Bing Xu\IEEEauthorrefmark{*}
    \IEEEauthorblockA{
        Department of Aeronautical and Aviation Engineering, The Hong Kong Polytechnic University, Hong Kong, China\\}
        Emails: sijia-franz.li@connect.polyu.hk, seergio.vicenzo@connect.polyu.hk, pbing.xu@polyu.edu.hk}
}

\maketitle

\begin{abstract}
In recent years, the fifth-generation (5G) new radio (NR) signals have emerged as a promising supplementary resource for urban navigation. However, a major challenge in utilizing 5G signals lies in their vulnerability to non-line-of-sight (NLoS) propagation effects, which are especially prevalent in urban street canyons. This paper applies the direct position estimation (DPE) method to 5G cellular signals to mitigate the NLoS bias as well as the multipath effects, thereby enabling precise localization in urbanized environments. The feasibility of applying the DPE method to NR positioning is analyzed, followed by a discussion of the tapped delay line (TDL) channel propagation model provided by the 3rd Generation Partnership Project (3GPP). The positioning performance is then evaluated through large-scale system-level simulations. The simulation results demonstrate that 5G DPE achieves satisfactory positioning accuracy in a 10 dB noisy channel, with an overall root mean square error (RMSE) constrained within 6 m. In addition, 5G DPE outperforms the observed time difference of arrival (OTDoA) method by 95.24\% in terms of positioning accuracy in an NLoS-dominated propagation environment.

\end{abstract}

\begin{IEEEkeywords}
5G positioning, maximum likelihood estimator, urban navigation
\end{IEEEkeywords}

\section{Introduction}
Precise localization is indispensable to a wide range of critical applications, including autonomous ground vehicles (AGVs) \cite{b1} and swarm intelligence \cite{b2}, both of which depend on accurate position information to ensure operational safety, efficiency, and reliability. It is particularly essential in densely populated urban areas, where the complexity of navigation and the existence of dynamic obstacles pose significant challenges. Among existing positioning technologies, the Global Navigation Satellite System (GNSS) is the most predominant method due to its global coverage and high accuracy in open environments. However, its performance deteriorates substantially in urbanized settings, where dense infrastructure and complex surroundings result in severe signal blockage, reflection, and multipath effect \cite{b3}.

The utilization of non-GNSS radio signals for positioning, navigation, and timing (PNT) has received significant attention in recent years and is increasingly regarded as a viable supplement when line-of-sight (LoS) GNSS observables are limited \cite{b4}. Among these, the fifth generation (5G) new radio (NR) cellular signals are of particular interest from both academia \cite{b5,b6} and industry \cite{b7,b8,b9}, owing to their high signal bandwidth, favorable geometric diversity, and extensive distribution density. The major challenge in exploiting 5G signals for positioning lies in their vulnerability to non-line-of-sight (NLoS) propagation effect, which is particularly common in urban street canyons \cite{b10}. The resulting NLoS bias introduces inevitable timing errors in ranging estimation, significantly compromising positioning accuracy \cite{b11}. In addition, the major NR positioning methods, such as multi-round trip time (multi-RTT), observed time difference of arrival (OTDoA), and uplink time difference of arrival (UL-TDoA), require stringent synchronization either between the user equipment (UE) and the base station (BS) or among multiple BSs \cite{b12}. Consequently, there is an urgent need for research into alternative algorithms that are less susceptible to the challenges posed by NLoS conditions and synchronization imperfection in NR positioning.

Direct position estimation (DPE) has been proposed as a more robust algorithm in terms of achieving precise positioning in harsh propagation scenarios, which utilizes an idea of solving the receiver position, velocity, and timing (PVT) directly via maximum likelihood (ML) in the navigation domain \cite{b13,b14,b15,b16,b17}. Initially proposed in \cite{b13}, DPE has been demonstrated to outperform traditional methods such as time of arrival (ToA), angle of arrival (AoA), and their combinations. The authors of \cite{b14} extended the concept to GNSS receivers and further generalized it to multi-antenna systems \cite{b15}. The theoretical analysis of DPE, which provided a mathematical perspective on its performance metrics, was carried out by deriving the Cramér-Rao bound (CRB) \cite{b16} and the Ziv-Zakai bound (ZZB) under both high and low signal-to-noise ratio (SNR) conditions, both of which demonstrated that DPE surpasses the traditional two-step (2SP) method in positioning accuracy under multipath environments.

Currently, DPE for 5G NR signals remains an unexplored area of research. The primary challenge for 5G DPE lies in the difference of the channel model for NR signals propagating in urban environments compared to that of GNSS reception. Focusing on applying the direct estimation method to NR positioning in dense urban area, this paper has the following contributions:
\begin{itemize}
    \item Both the LoS and NLoS multipath channel models for 5G NR signal propagation are analyzed. This paper demonstrates that the DPE method is particularly effective in mitigating NLoS bias with multipath, resulting in superior positioning performance in NLoS-dominated scenarios.
    \item The mathematical model for 5G DPE is derived in this paper, followed by large-scale system-level simulations with an urbanized setting for algorithm validation.
\end{itemize}

\section{DPE for 5G NR Positioning}
This section first examines the multipath channel characteristics of 5G NR under both NLoS and LoS conditions, highlighting the necessity of deploying the DPE method for NR-based positioning. Following this, a detailed mathematical description of the 5G DPE model is presented.

\subsection{5G NR Multipath Propagation Channel}
NR adopts the tapped delay line (TDL) channel model for non-multiple-input-multiple-output (non-MIMO) evaluations, wherein each tap represents a multipath component characterized by a fixed delay, a Rayleigh-distributed power, and a classical Jakes Doppler spectrum \cite{b18}. The TDL propagation channel model is defined as
\begin{equation}
    h(\tau,t) = \sum_{l=0}^{L-1}h_l(t)\delta(\tau-\tau_l)
\end{equation}
where $\left\{ h_l(t) \right\}_{l=0}^{L-1}$ are the wide-sense stationary (WSS) narrow-band complex Gaussian processes, $\left\{\tau_l\right\}_{l=0}^{L-1}$ are the tap delays, and $L$ is the number of channel taps.

\begin{figure}[h]
    \centering
    \includegraphics[width=1\linewidth]{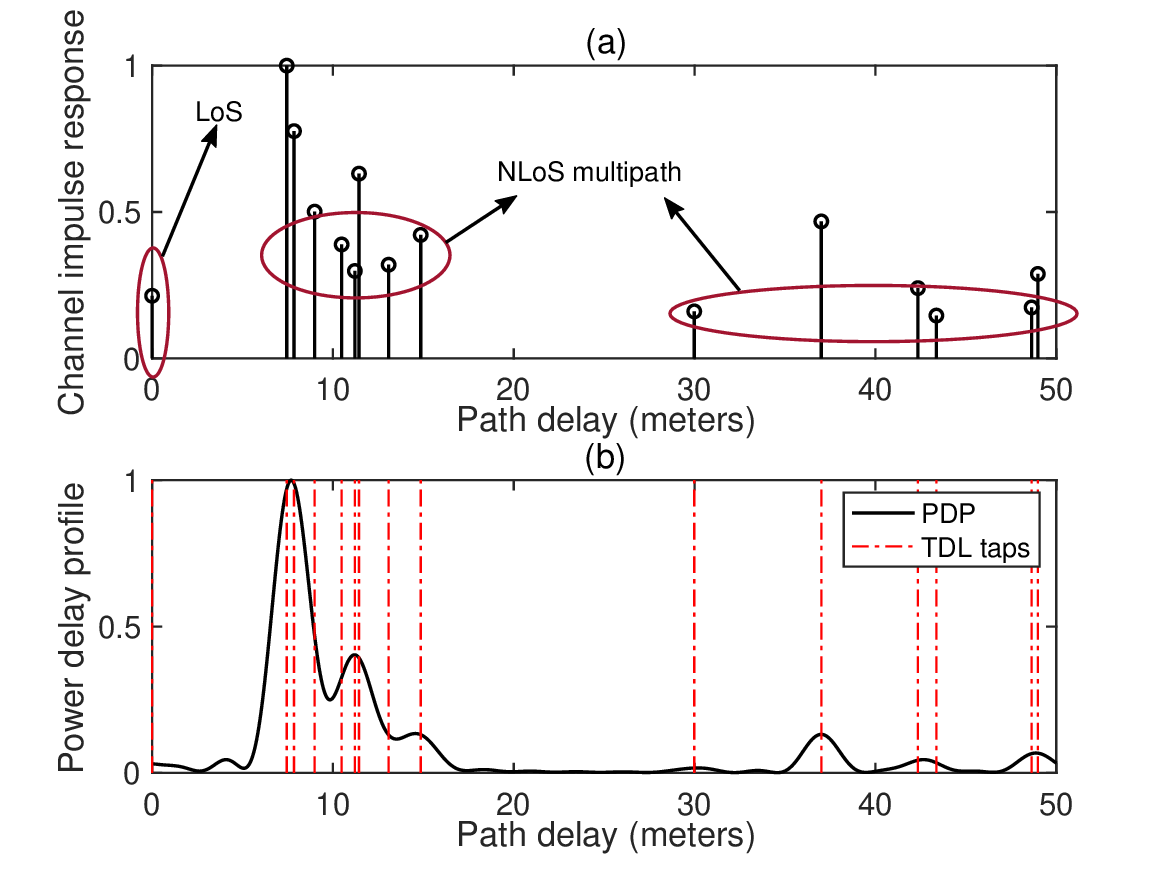}
    \caption{TDL-A channel model for NLoS condition generated with a desired RMS delay spread of 65 ns. (a) is the channel impulse response of the TDL model, (b) is the power delay profile quantifying the channel characteristics.}
    \label{tdl-a}
\end{figure}

Distinct TDL profiles are specified for different deployment scenarios by the 3rd Generation Partnership Project (3GPP), including Indoor (InH), Urban Microcell (UMi), and Urban Macrocell (UMa) environments \cite{b19}. Specifically, three TDL models, namely TDL-A, TDL-B, and TDL-C, are designed to characterize NLoS channel conditions, whereas TDL-D and TDL-E are employed to represent LoS propagation scenarios. The normalized tap delays of each TDL model can be scaled to achieve a desired root mean square (RMS) delay spread
\begin{equation}
    \tau_l^{\text{actual}} = \tau_l^{\text{normalized}} \times \text{DS}_{\text{desired}}
\end{equation}
where $ \tau_l^{\text{actual}}$ represents the actual delay for tap $l$, and $\tau_l^{\text{normalized}}$ denotes the normalized tap delay as defined in a specific TDL model. $\text{DS}_{\text{desired}}$ corresponds to the target RMS delay spread, which quantifies the temporal dispersion caused by multipath propagation and is defined as the time difference between the earliest and latest arrivals of signal components. 
\begin{figure}[t]
    \centering
    \includegraphics[width=1\linewidth]{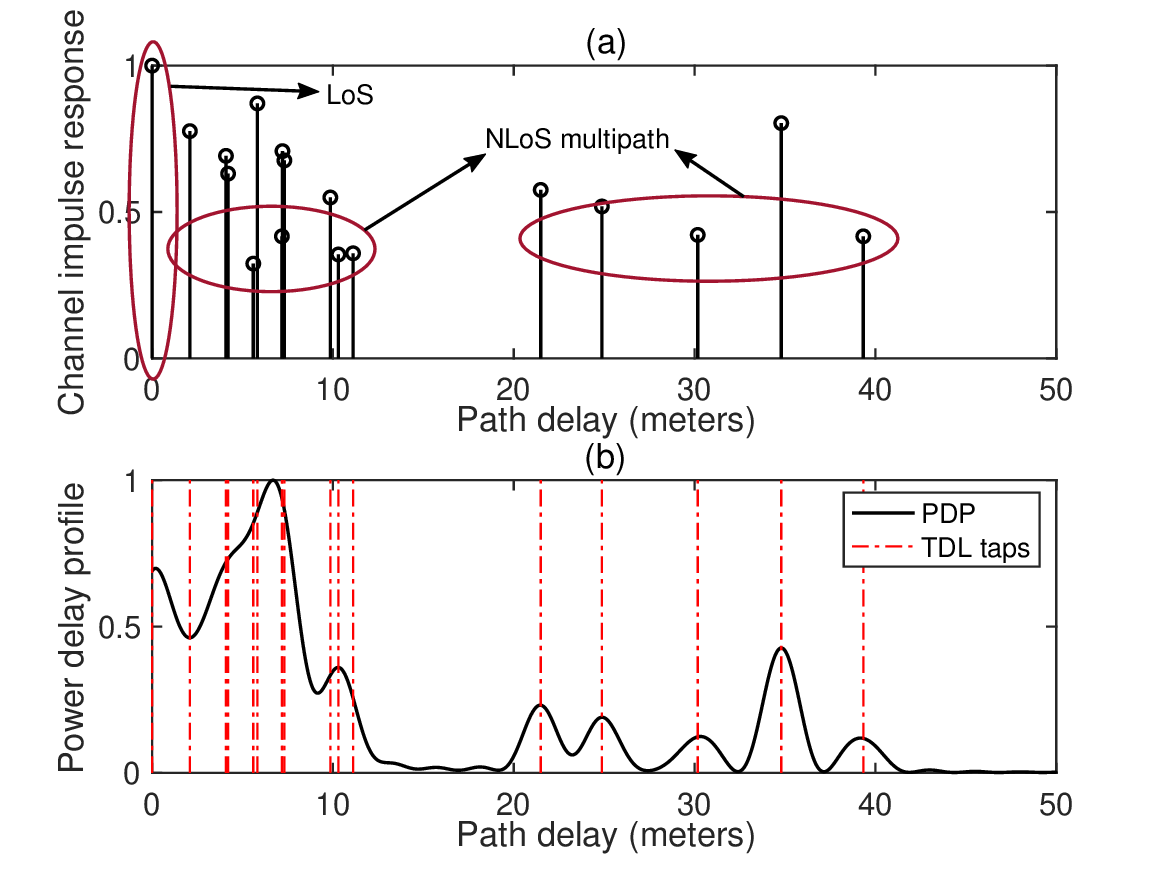}
    \caption{TDL-B channel model for NLoS condition generated with a desired RMS delay spread of 65 ns. (a) is the channel impulse response of the TDL model, (b) is the power delay profile quantifying the channel characteristics.}
    \label{tdl-b}
\end{figure}
\begin{figure}[b]
    \centering
    \includegraphics[width=1\linewidth]{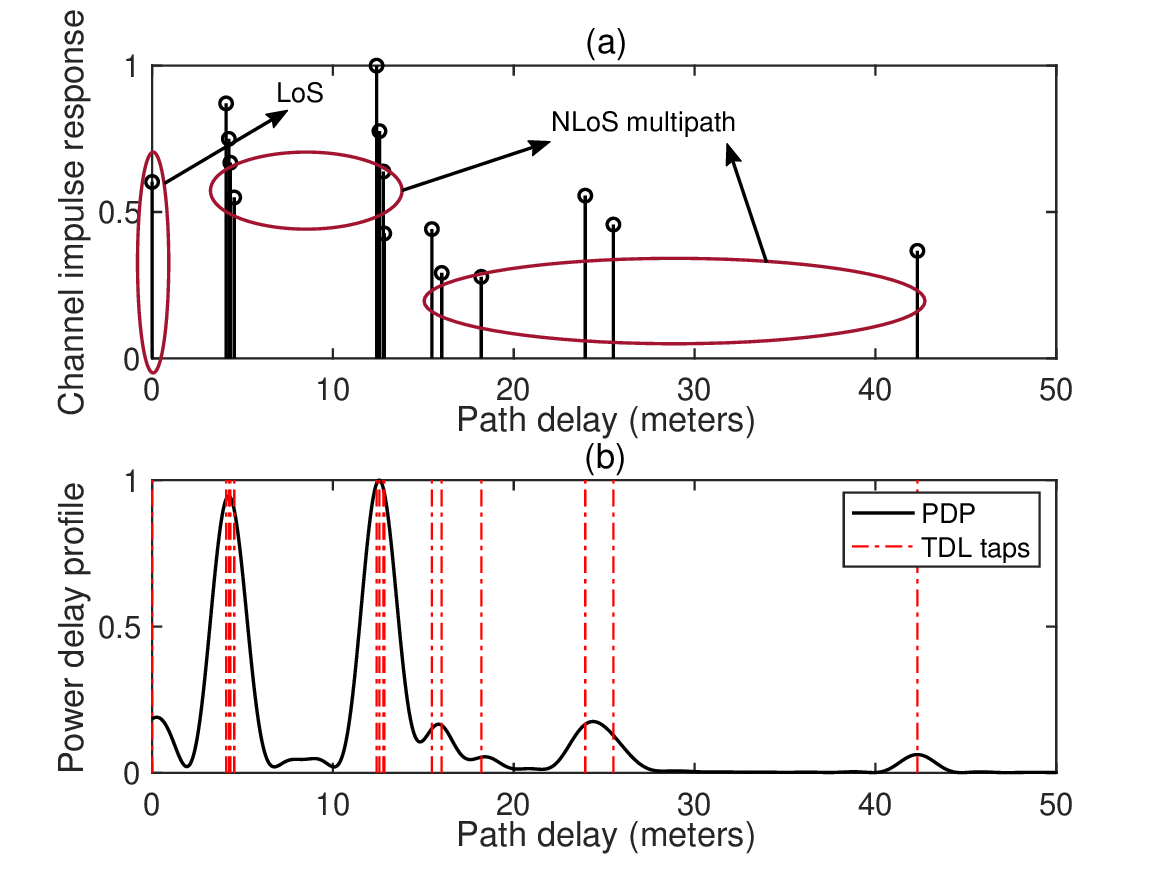}
    \caption{TDL-C channel model for NLoS condition generated with a desired RMS delay spread of 65 ns. (a) is the channel impulse response of the TDL model, (b) is the power delay profile quantifying the channel characteristics.}
    \label{tdl-c}
\end{figure}

Fig.~\ref{tdl-a}-\ref{tdl-c} respectively illustrate the 3GPP NLoS channel models for TDL-A, TDL-B, and TDL-C in terms of both the channel impulse response (CIR) and the resulting normalized power delay profile (PDP) for each model, generated with a desired RMS delay spread of 65 ns, specified for UMi street canyon environment with a short-delay profile. It can be observed that for NLoS propagation scenario, there is no multipath component occupying a significantly higher power than other reflected paths, while it still has a first arriving path but has a relatively low power, indicating that the LoS signal is partially blocked. The joint NLoS and multipath impact posed on the signal propagated in the channel inevitably leads to a timing error around 7-12 m, which can be even larger for UMi and UMa use cases with longer delay profiles \cite{b20}.

\begin{figure}[htbp]
    \centering
    \includegraphics[width=1\linewidth]{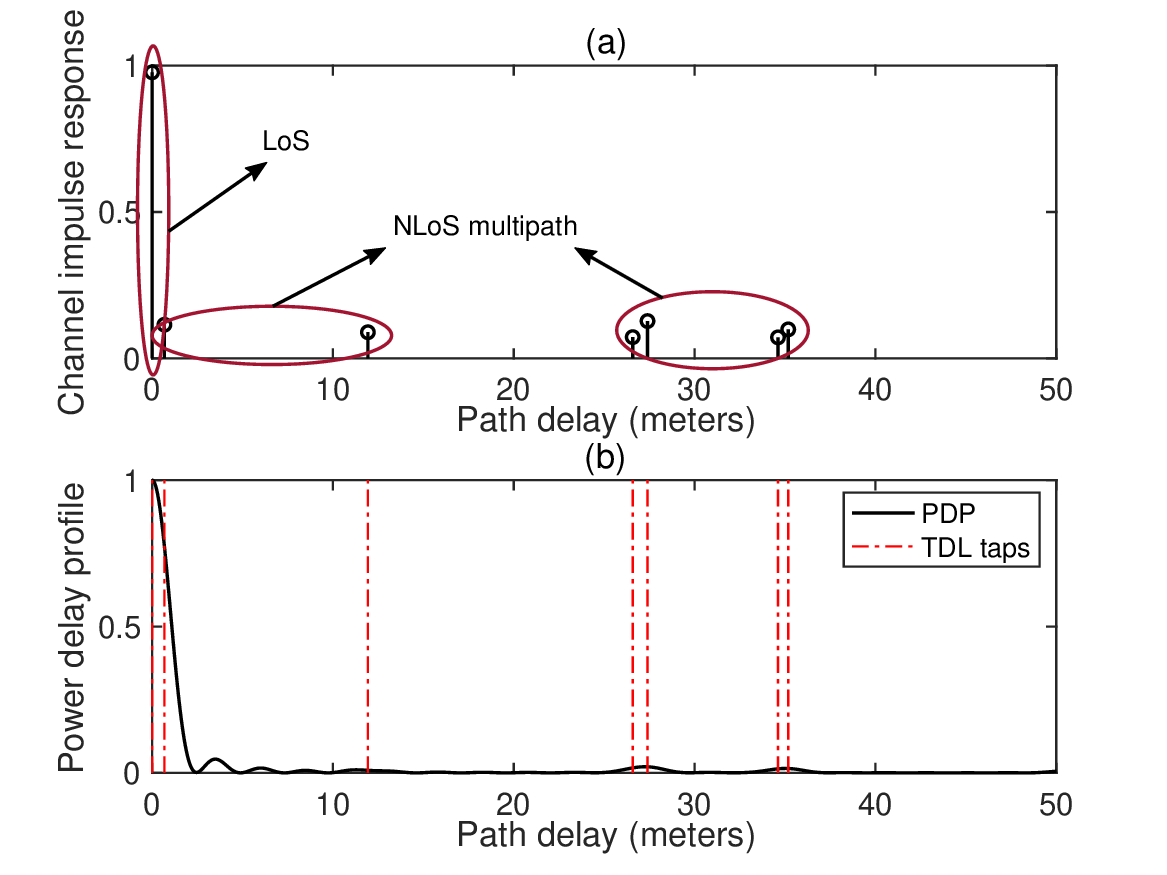}
    \caption{TDL-D channel model for LoS condition generated with a desired RMS delay spread of 65 ns. (a) is the channel impulse response of the TDL model, (b) is the power delay profile quantifying the channel characteristics.}
    \label{tdl-d}
\end{figure}

\begin{figure}[htbp]
    \centering
    \includegraphics[width=1\linewidth]{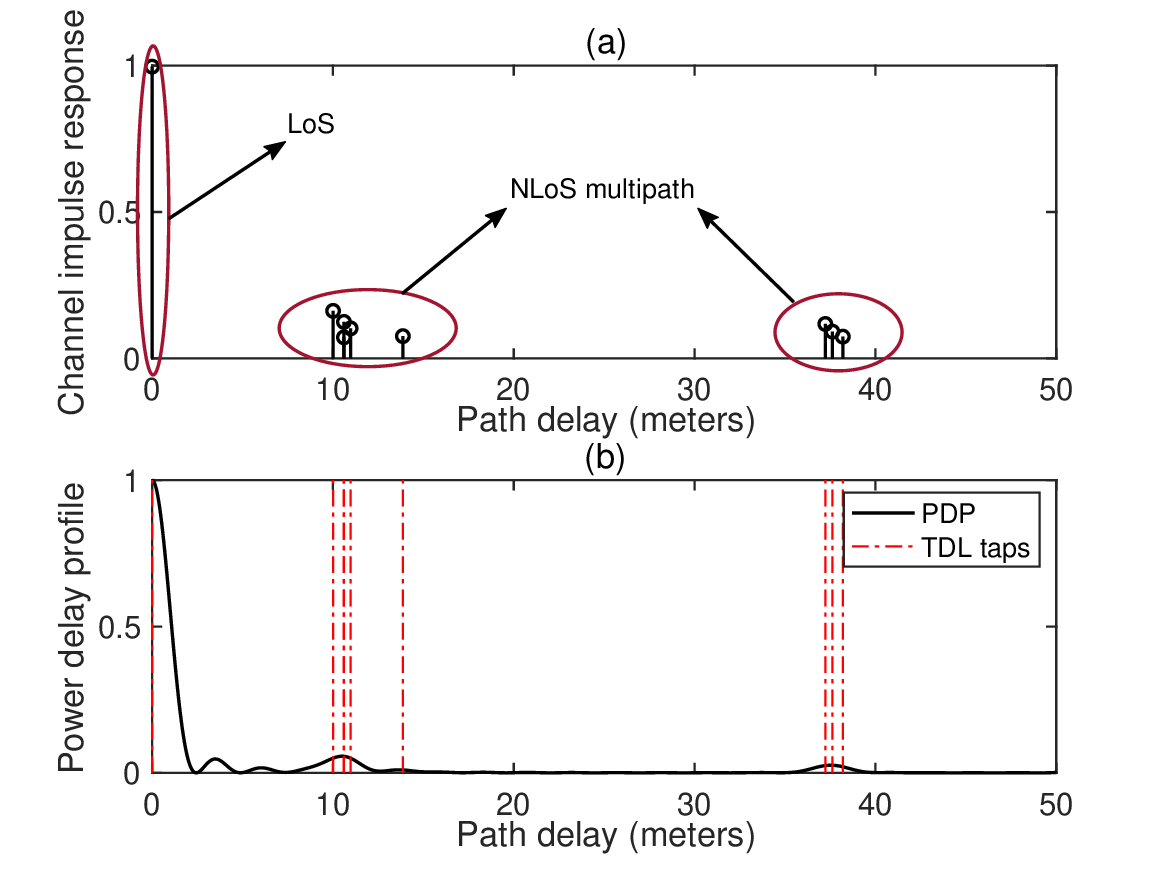}
    \caption{TDL-E channel model for LoS condition generated with a desired RMS delay spread of 65 ns. (a) is the channel impulse response of the TDL model, (b) is the power delay profile quantifying the channel characteristics.}
    \label{tdl-e}
\end{figure}

TDL-D and TDL-E are the channel models for the LoS conditions. As indicated in Fig.~\ref{tdl-d}-\ref{tdl-e}, 3GPP LoS channels also consist of multipath, but the first arriving path dominates the multipath and carries the strongest power, which is outstandingly higher than the summation of all power located on other multipath components. Therefore, for LoS propagation channel, the accuracy of timing estimation is predominantly influenced by the intrinsic properties of the transmitted signal, such as its bandwidth, rather than by multipath effects. For cellular signals with limited bandwidth, such as 4G long-term evolution (LTE) signals, where the system bandwidth ranges from 1.4 MHz to 20 MHz, the receiver ranging performance degrades in LoS propagation scenarios due to the presence of multipath effects \cite{b21,b22}. This degradation occurs because the multipath resolution is significantly smaller than the chip duration of the signal, leading to a distortion in the shape of the auto-correlation function (ACF). As a result, the multipath interference introduces timing errors, adversely affecting the accuracy of ToA measurements. This paper focuses on full bandwidth 5G signals, hence the impact of multipath on LoS propagation scenarios is negligible. 




\subsection{Mathematical Model for 5G DPE}
The channel characteristics of NR suggest that the DPE method is particularly effective at correcting NLoS bias, as it directly aggregates all the first-arriving path components from each contributing BS, thereby improving the accuracy of position estimation in NLoS conditions. Consider a single-input-single-output (SISO) cyclic prefix orthogonal frequency division multiplexing (CP-OFDM) system consisting of $N$ subcarriers for NR downlink data transmission, the baseband signal transmitted at the $m$th BS can be represented as
\begin{equation}
    x_m(t) = \frac{1}{\sqrt{N}} \sum_{k=0}^{N-1} d_{m,k} e^{j2\pi k\Delta f(t-T_{\rm{CP}} -iT_{\rm{symb}})}
\end{equation}
for $iT_{\rm{symb}} \leq t < (i+1)T_{\rm{symb}}$, where $i$ is the OFDM symbol index. $T_{\rm{symb}} = T_{\rm{CP}} + T_{\rm{u}}$ represents the whole OFDM symbol duration, including the actual data length $T_{\rm{u}}$ and the duration of cyclic prefix (CP) $T_{\rm{CP}}$. The introduction of CP added at the head of each symbol is aimed at eliminating the inter-symbol interference (ISI) caused by multipath. $d_{m,k}$ is the modulated data for transmission, $\Delta f = 1/T_{\rm{u}}$ is the subcarrier spacing (SCS). Without loss of generality, the following assumptions are adopted in this paper:
\begin{itemize}
    \item The dedicated 5G positioning reference signal (PRS) occupying the full system bandwidth is utilized and frequency resource pattern is not a primary concern, which indicates that the quadrature phase shift keying (QPSK) modulated pilot signals with normalized power allocation are used for downlink transmission and it is known by the UE.
    \item The system is considered to be static regardless of the UE mobility. Hence the Doppler effect is assumed to be negligible. In addition, both the carrier frequency offset (CFO) and phase noise (PHN) are assumed to be perfectly estimated and pre-compensated before the positioning procedure.
    \item The wireless channel is assumed to be quasi-stationary due to the assumption that the CP duration is longer than the channel maximum excess delay. Hence, the channel remains the same within one OFDM symbol duration but varies from symbol to symbol, i.e., $h_l(t) = h_l$, for $iT_{\rm{symb}} \leq t < (i+1)T_{\rm{symb}}$.
\end{itemize}

At the UE side, the received OFDM signal followed by the CP removal is given by
\begin{align}
    \begin{split}
         y_m(t) &= \sum_{l=0}^{L-1}h_{m,l}x_m(t-\tau _{m,l}(\mathbf{p})) + \eta_m(t)\\
         &= \frac{1}{\sqrt{N}}\sum_{l=0}^{L-1}h_{m,l}\sum_{k=0}^{N-1}d_{m,k}e^{j2\pi k\Delta f(t-\tau_{m,l}(\mathbf{p}))}+\eta_m(t)
    \end{split}
\end{align}
where $\left\{h_{m,l}\right\}_{l=0}^{L-1}$ are the complex channel coefficients for the signal propagating from the $m$th BS to the UE in an $L$-tap multipath channel. $\left\{\tau_{m,l}(\mathbf{p})\right\}_{l=0}^{L-1}$ are the multipath delays, which depend on the UE positioning information $\mathbf{p} = [\delta x, \delta y, \delta z, \delta t]^\top$ with respect to the $m$th BS. $\eta_m(t)$ is the complex additive white Gaussian noise (AWGN) with zero mean and known variance $\sigma_{\rm{AWGN}}^2$. The received OFDM signal is sampled at a sampling interval $T_{\rm{s}} = T_{\rm{u}}/N$. For signals occupying the full system bandwidth, the number of subcarriers $N$ is equivalent to the discrete Fourier transform (DFT) points. The normalized $n$th discrete time received sample can be expressed as
\begin{align}\label{eq3}
\begin{split}
y_m(nT_{\rm{s}}) &= h_{m,v}\sum_{k=0}^{N-1}d_{m,k}e^{\frac{j2\pi kn}{N}}e^{-j2\pi k\Delta f\tau_m{(\mathbf{p})}}\\
&\quad +\sum_{l=0,l\neq v}^{L-1}h_{m,l}\sum_{k=0}^{N-1}d_{m,k}e^{\frac{j2\pi k(n-\tau_l)}{N}}
\end{split}
\end{align}
for $n = 0,\cdots, N-1$, where $v$ is the LoS index among the multipath, and $\tau_m(\mathbf{p})$ is the ToA of the $m$th BS in the function of $\mathbf{p}$. The received signal $\mathbf{y}_m \in \mathbb{C}^{N\times1}$ can be hence written in the vector form as
\begin{equation}\label{eq4}
    \mathbf{y}_m = \mathbf{F}^H\mathbf{D}_m\mathbf{\Gamma}_m(\mathbf{p})\mathbf{F}_L\mathbf{h}_m + \boldsymbol{\eta}_m
\end{equation}
where 
\begin{itemize}
    \item $\mathbf{F} \in \mathbb{C}^{N\times N}$ is the DFT matrix, with each entity being $[\mathbf{F}]_{\iota,\kappa} = (1/\sqrt{N})e^{j2\pi \iota \kappa /N}$, for $\iota, \kappa = 0,\cdots,N-1$.
    \item $\mathbf{D}_m = \text{diag}[d_{m,0}, \cdots, d_{m,N-1}]^\top \in \mathbb{C}^{N \times N}$ is the modulated PRS data with normalized power and to be transmitted by the $m$th BS.
   \item $\mathbf{\Gamma}_m(\mathbf{p}) = 
\text{diag}\big[1, e^{-j2\pi \Delta f\tau_m(\mathbf{p})}, \\
\hphantom{1}\hphantom{1}\hphantom{1}\hphantom{1}\hphantom{1}\hphantom{1}\hphantom{1}\hphantom{1}\hphantom{1}\hphantom{1}\hphantom{1}\hphantom{1}\hphantom{1}\hphantom{1}\hphantom{1}\cdots, e^{-j2\pi (N-1)\Delta f\tau_m(\mathbf{p})}\big]^\top \in \mathbb{C}^{N \times N}$ is the ToA matrix containing UE position information.
\item $\mathbf{F}_L = \mathbf{F}(1:N,1:L)\in \mathbb{C}^{N\times L}$ is the DFT matrix.
\item $\mathbf{h}_m = [h_{m,0},\cdots,h_{m,L-1}]^\top \in \mathbb{C}^{L\times 1}$ is the CIR vector for the signal propagating from the $m$th BS to the UE.
\item $\boldsymbol{\eta}_m = [\eta_0,\cdots,\eta_{N-1}] \in \mathbb{C}^{N\times 1 }$ is the noise vector.
\end{itemize}
The received signal model exposed in (\ref{eq4}) is a function of UE positioning information $\mathbf{p}$, which consists of the 3D coordinates and the clock bias. The objective is to obtain the ML estimate of the UE position directly from the received signal model. Considering a $M$-BS downlink positioning scenario in an AWGN channel, the ML solution is an equivalent to the least squares (LS) estimation, followed by minimizing the cost function summed over all observables from $M$ BSs
\begin{align}\label{eq5}
\begin{split}
     \Lambda (\mathbf{p}) &= \sum_{m=1}^{M}{\left\Vert  \mathbf{y}_m - \mathbf{F}^H \mathbf{D}_m\mathbf{\Gamma}_m(\mathbf{p})\mathbf{F}_L\mathbf{h}_m \right\Vert ^{2}}\\
    &= \sum_{m=1}^{M}\mathbf{y}_m^{H}\mathbf{y}_m - 2\Re\left\{\mathbf{y}_m^{H}\mathbf{F}^H\mathbf{D}_m\mathbf{\Gamma}_m(\mathbf{p})\mathbf{F}_L\mathbf{h}_m\right\}\\
    &\phantom{=}\phantom{=}\phantom{=}+ \mathbf{h}_m^{H}\mathbf{F}_{L}^{H}\mathbf{\Gamma}_m^{H}(\mathbf{p})\mathbf{D}_m^{H}\mathbf{F}\mathbf{F}^H\mathbf{D}_m\mathbf{\Gamma}_m(\mathbf{p})\mathbf{F}_L\mathbf{h}_m
\end{split}
\end{align}
where $\Re\left\{\cdot\right\}$ denotes the real parts of a complex quantity. The estimate of $\mathbf{h}_m$ that minimizes the cost function is given by
\begin{equation}\label{eq6}
    \hat{\mathbf{h}}_m = \underbrace{\left(\mathbf{F}_{L}^{H}\mathbf{\Gamma}_m^{H}(\mathbf{p})\mathbf{D}_m^{H}\mathbf{D}_m\mathbf{\Gamma}_m(\mathbf{p})\mathbf{F}_L\right)^{-1}}_{\textstyle \mathbf{I}_L}\mathbf{F}_L^H\mathbf{\Gamma}_m^{H}(\mathbf{p})\mathbf{D}_m^{H}\mathbf{F}\mathbf{y}_m
\end{equation}
Substituting (\ref{eq6}) in (\ref{eq5}) we have
\begin{align}
    \begin{split}
       \Lambda (\mathbf{p}) &= \sum_{m=1}^{M}\mathbf{y}_m^{H}\mathbf{y}_m - \\
       &\phantom{=}\phantom{=}\phantom{=}\phantom{=}\phantom{=}\mathbf{y}_m^{H}\mathbf{F}^H\mathbf{D}_m\mathbf{\Gamma}_m(\mathbf{p})\mathbf{F}_L\mathbf{F}_L^H\mathbf{\Gamma}_m^{H}(\mathbf{p})\mathbf{D}_m^{H}\mathbf{F}\mathbf{y}_m
    \end{split}
\end{align}
Thus, the ML estimates of $\mathbf{p}$ is given by maximizing the resulting negative cost function $\tilde{\Lambda} (\mathbf{p})$
\begin{align}\label{eq_costFunc}
    \begin{split}
        \hat{\mathbf{p}}_{\rm{ML}} &= \mathop{\arg\max}_{\textstyle\mathbf{p}} \left\{\tilde{\Lambda} (\mathbf{p})\right\}\\
        &= \mathop{\arg\max}_{\textstyle\mathbf{p}} \left\{\sum_{m=1}^{M}\mathbf{y}_m^{H}\mathbf{S}_m(\mathbf{p})\mathbf{S}_m^{H}(\mathbf{p})\mathbf{y}_m\right\}
    \end{split}
\end{align}

\begin{figure}[h]
    \centering
    \includegraphics[width=\linewidth]{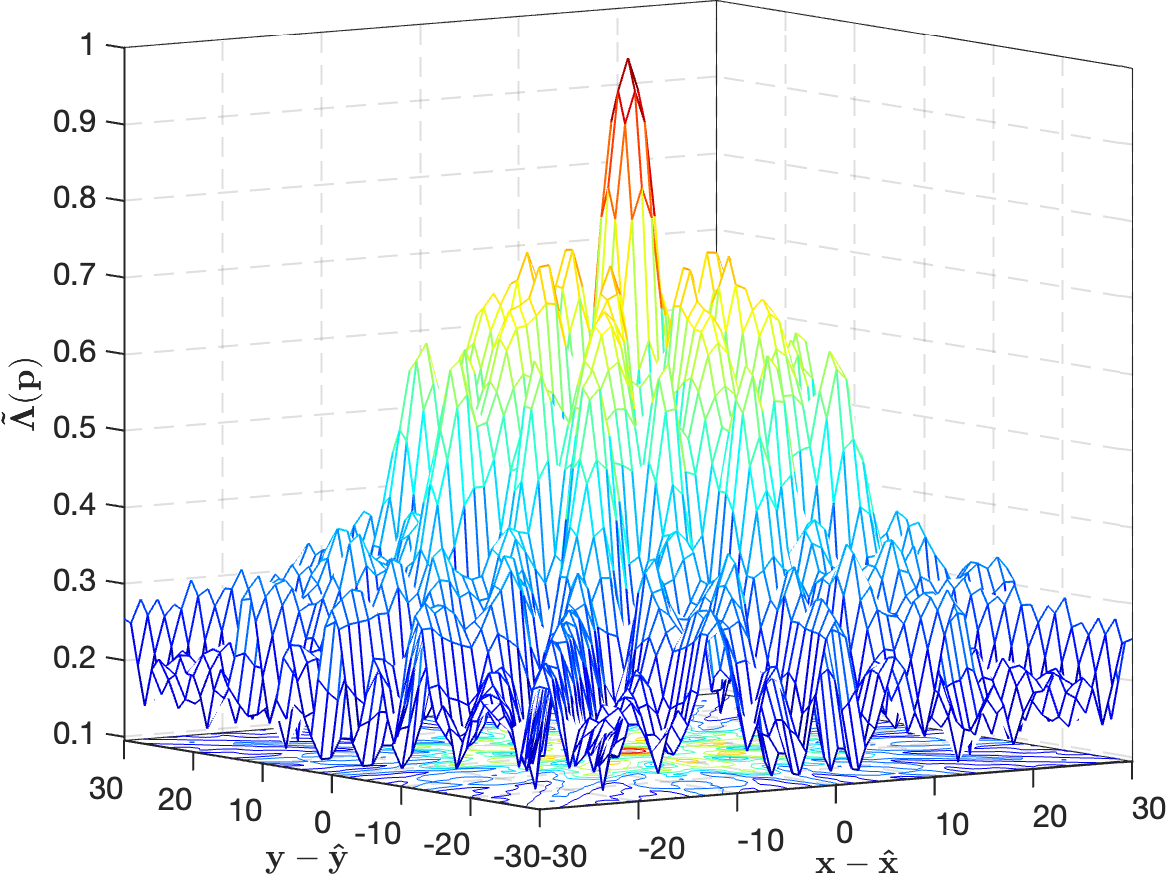}
    \caption{Cost function composed of signals from 8 BSs, with each signal propagating through an equally spaced, 3-tapped NLoS channel.}
    \label{dpe}
\end{figure}
where $\mathbf{S}_m(\mathbf{p}) = \mathbf{F}^H\mathbf{D}_m\mathbf{\Gamma}_m(\mathbf{p})\mathbf{F}_L$ is the basis function matrix, representing the delayed PRS time domain signal convolving with the multipath channel. Fig.~\ref{dpe} illustrates the cost function in (\ref{eq_costFunc}), which is is derived from signals received from 8 BSs, and obtained for the evaluation of 2D position errors. For simplification, each transmitted signal undergoes propagation through a 3-tap NLoS channel, where the amplitude of the LoS signal and the second multipath are 3 dB and 1 dB lower, respectively, than that of the first multipath. Specifically, the amplitudes of the CIR are $|h_1| = 0.7079$, $|h_2| = 1$ and $|h_3| = 0.8913$, and the channel is assumed to be the same for all BSs. The full bandwidth downlink PRS is utilized with $N = 4096$. Each tap is spaced equally by one sample, corresponding to a distance of 2.44 m between adjacent taps. As indicated in Fig.~\ref{dpe}, the maximization of the cost function in (\ref{eq_costFunc}) leads to a direct estimation of the user position $\mathbf{P} = [x,y]^\top$. For the sake of clarification, the estimation for altitude and clock bias is not involved in this example.

\section{System-Level Simulation in Urban Environment}
\begin{figure}[h]
    \centering
    \includegraphics[width=1\linewidth]{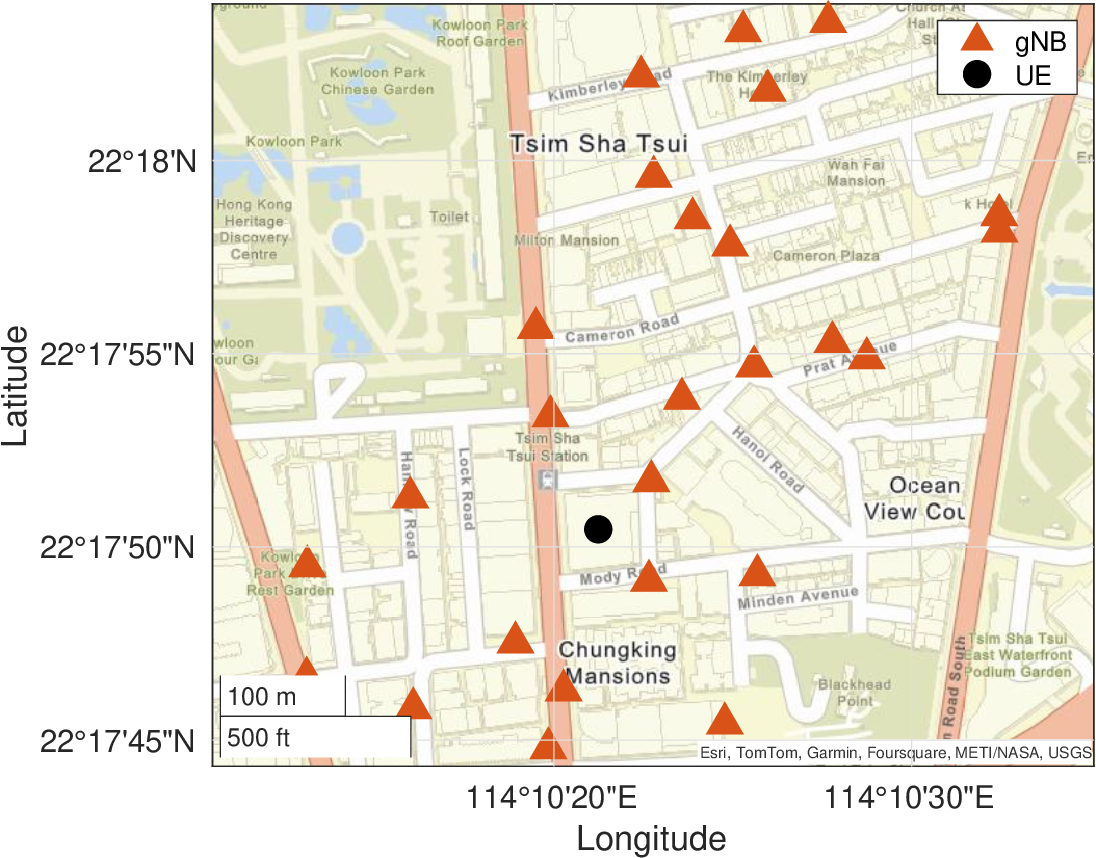}
    \caption{Realistic cellular deployment established in a densely populated urban environment in Tsim Sha Tsui, consisting of 26 BSs distributed across a 227,360 $\text{m}^2$ region.}
    \label{map}
\end{figure}

This section presents the system-level simulation for proposed 5G DPE. The simulation is conducted within a realistic cellular deployment scenario, set in the densely populated urban area of Tsim Sha Tsui, Hong Kong. A total of 26 BSs are distributed across a 227,360 $\text{m}^2$ square region. For each candidate BS, the transmission parameters are strictly defined in accordance with specifications in \cite{b18}. The cellular deployment is modeled as an UMi street canyon scenario, with each BS situated at an average height of 10 m and transmitting with a power of 24 dBm. The details of the simulated network deployment is illustrated in Fig.~\ref{map}.

The NR downlink transmission for each BS is simulated using a dedicated PRS occupying a bandwidth of 100 MHz. It is assumed that all BSs are perfectly synchronized, while the synchronization error between the UE and the BS is considered. This error is modeled to have a truncated Gaussian distribution with a standard deviation of 5 ns followed by a coarse downlink and uplink synchronization in the random access (RA) procedure \cite{b24}. The timing offset induced by the synchronization error is thus modeled as a constant phase shift, linearly applied to the OFDM symbols in the frequency domain of the received signal. The LoS condition for each BS is determined based on the LoS probability described in \cite{b18}, which is expressed as
\begin{equation}
\text{Pr}_{\rm{LoS}} =
\begin{cases}
\frac{18}{D} +  \left(1 - \frac{18}{D}\right)e^{-\frac{D}{36}}  , &  D > 18\\
1, &  D \leq 18
\end{cases}
\end{equation}
where $D$ is the 2D geometry distance between the BS and UE. The propagation channel for each transmission of the candidate BS is classified as either LoS or NLoS based on the distance-dependent LoS condition. The TDL-D and TDL-C channel profiles are then applied for the LoS and NLoS cases, respectively, with an overall delay spread of 65 ns.
\begin{figure}[h]
    \centering
    \includegraphics[width=\linewidth]{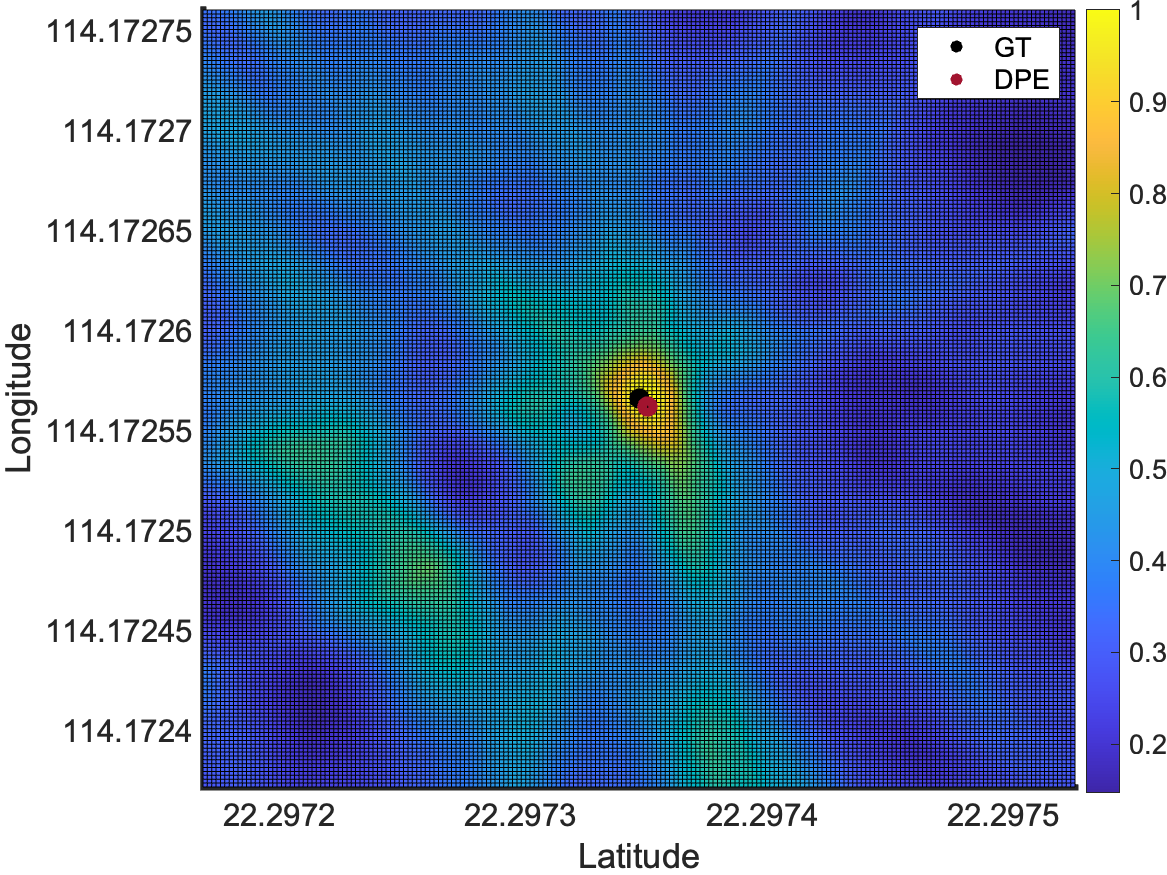}
    \caption{2D correlogram for 5G DPE under the aforementioned urbanized positioning scenario.}
    \label{dpe_exp}
\end{figure}

At the UE side, a divide-and-conquer-based DPE algorithm is deployed, which aims at generating a correlogram by correlating the received signal from a single BS with the local replica generated at each candidate point around the actual UE position \cite{b25,b26}. The resulting correlogram is obtained by summing the contributions from all BSs. The UE position is then estimated by optimizing the summed correlogram, which effectively corresponds to find the peak point location. The correlogram for 5G DPE under the aforementioned urbanized positioning scenario is given in Fig.~\ref{dpe_exp}, where the result indicates that the DPE result has little bias to the ground truth (GT) of the UE.

\section{Simulation Results}
This section presents the system-level simulation results for evaluating the performance of the proposed 5G DPE, using the simulation environment outlined in Section III. For each simulated scenario, a Monte Carlo simulation is conducted with 5000 trials to ensure statistically significant results. The LoS propagation cases account for only 17\% of the total simulations, while 83\% of the cases correspond to NLoS channels. This indicates that the results are primarily derived from an NLoS-dominant propagation scenario. The clock bias is set with a deviation of 5 ns for the simulation of imperfect synchronization. To provide a comprehensive analysis, we first present the positioning accuracy of the proposed 5G DPE under varying noise levels and for different numbers of BS contributing to the positioning. Subsequently, a comparative analysis is conducted to jointly evaluate the positioning performance of 5G DPE and OTDoA. The detailed simulation parameters are provided and summarized in Table~\ref{tab}.
\begin{table}[t]
\renewcommand{\arraystretch}{1.5}
\caption{Simulation Parameters}
\begin{center}
\begin{tabularx}{8.5cm}{p{5.5cm}|m{2.2cm}}\hline\hline
\centering \textbf{Description} &\centering \textbf{Value}   \tabularnewline\hline

\centering BS height (m) &\centering 10 \tabularnewline\hline
\centering UE height (m)  &\centering 2 \tabularnewline\hline
\centering Delay spread (ns) &\centering 65 \tabularnewline\hline
\centering Mobile speed (m/s) &\centering 0 \tabularnewline\hline
\centering Rician K-factor (dB) &\centering 13.3 \tabularnewline\hline
\centering Center frequency (GHz) &\centering 3 \tabularnewline\hline
\centering Transmission power (dBm) &\centering 24 \tabularnewline\hline
\centering Number of transmit antennas &\centering 1\tabularnewline\hline
\centering Number of receive antennas &\centering 1 \tabularnewline\hline
 \centering PRS frequency reuse factor &\centering 6 \tabularnewline\hline
 \centering Subcarrier spacing (kHz) &\centering 30 \tabularnewline\hline
 \centering System bandwidth (MHz) &\centering 100  \tabularnewline\hline
 \centering Clock bias deviation (ns) &\centering 5\tabularnewline\hline
 \centering LoS channel profile &\centering TDL-C \tabularnewline\hline
  \centering NLoS channel profile &\centering TDL-D \tabularnewline\hline
\hline
\end{tabularx}
\label{tab}
\end{center}
\end{table}

\subsection{5G DPE Positioning Accuracy}
\begin{figure}[t]
    \centering
    \includegraphics[width=\linewidth]{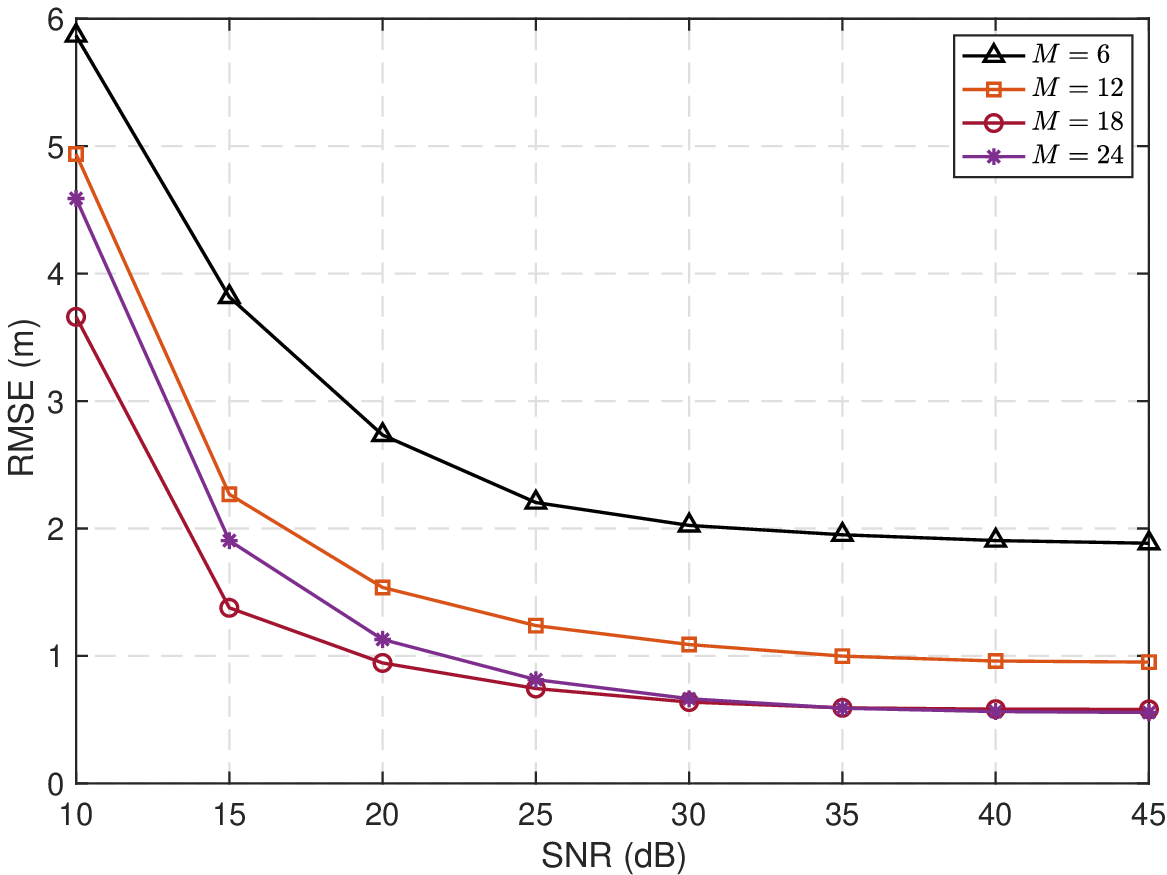}
    \caption{5G DPE positioning accuracy as a function of SNR for different numbers of BS contributing to the simulation.}
    \label{rmse}
\end{figure}

The 5G DPE positioning accuracy is evaluated using the root mean square error (RMSE), and its performance is examined under varying signal-to-noise ratio (SNR) levels, which represent the statistics of the noisy channel. Fig.~\ref{rmse} illustrates the positioning accuracy as a function of SNR for different numbers of BS contributing to the simulation. It can be observed that the 5G DPE generally achieves satisfactory positioning performance, with an overall RMSE constrained within 6 m, even in the worst simulation scenario where only 6 candidate BSs are used and the SNR is as low as 10 dB. As the number of available positioning sources increases, a sub-meter positioning accuracy can be achieved, particularly when the propagation environment provides a high signal quality. It is worth noticing that when the channel condition is not satisfactory enough, i.e., SNR below 20 dB, increasing the number of candidate BSs may introduce extra noise to the correlogram, an outcome resulting in an increase in the positioning error. Additionally, as indicated by the curves for $M=18$ and $M=24$, there is a theoretical limit to the performance of 5G DPE. Beyond a certain point, increasing the number of BS contributing to the positioning does not yield further improvements in accuracy and only adds unnecessary computational complexity.

\subsection{Comparative Analysis with OTDoA}
In this subsection, the performance of the proposed 5G DPE is further analyzed by comparing with the convention OTDoA method, which is frequently used for NR downlink positioning.

\begin{figure}[h]
    \centering
    \includegraphics[width=\linewidth]{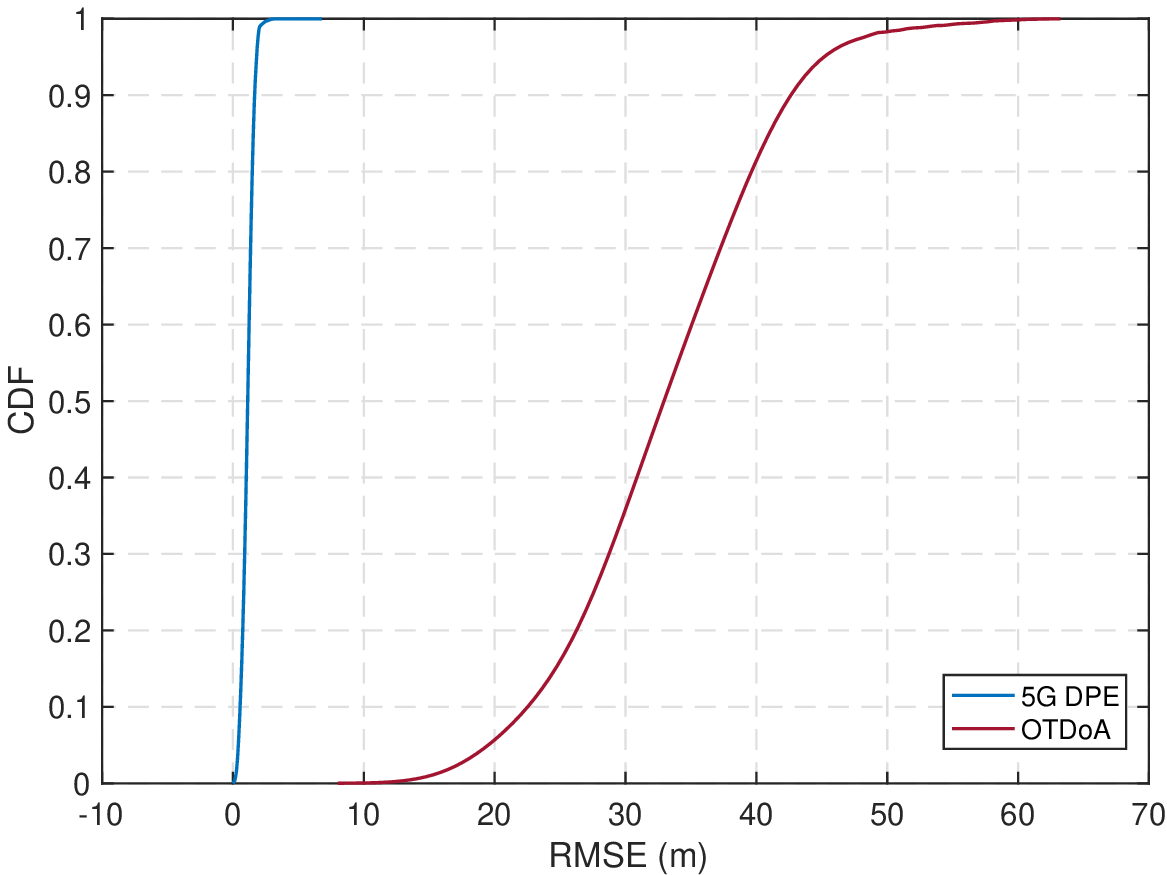}
    \caption{Positioning accuracy of the proposed 5G DPE and the OTDoA methods, evaluated by using four candidate BSs nearest to the UE position.}
    \label{cdf}
\end{figure}

Both the 5G DPE and OTDoA simulations utilize four candidate BSs for positioning, selected based on their minimum proximity to the UE within the specified positioning environment. The LoS conditions for each BS are determined according to their relative distance to the UE. Specifically, two of the four BSs experience LoS propagation, while the remaining two are affected by NLoS conditions. The SNR level for each transmission is calculated based on the path loss model of corresponding propagation channel for UMi street canyon in \cite{b18}. Fig.~\ref{cdf} presents the positioning accuracy of the proposed 5G DPE and the OTDoA methods respectively, where the algorithm performance is evaluated using the cumulative distribution function (CDF) of the RMSE. It can be observed that the RMSE obtained by 5G DPE can be constrained within 7 m for 99\% of the use cases, with a 90\% probability of achieving an RMSE within 2 m. In contrast, the conventional OTDoA method results in a 42-meter RMSE for 90\% of the use cases, which demonstrates that the proposed 5G DPE outperforms OTDoA by 95.24\%. This result indicates that the DPE method is more effective in correcting the NLoS biases introduced by the propagation channel, as the performance of the OTDoA method heavily relies on the accuracy of the ranging measurements. In an NLoS-dominated positioning scenario, the ranging measurements are prone to inaccuracies due to the presence of NLoS biases, as illustrated and discussed in Section II.

\section{Conclusion}
In this paper, we explore the application of the DPE method to 5G NR cellular signals within an urbanized positioning environment. A comprehensive analysis of the 3GPP NR multipath propagation channel model for both LoS and NLoS conditions is provided, demonstrating that the DPE method is particularly effective at mitigating NLoS biases, leading to enhanced positioning performance in NLoS-dominated scenarios. Focusing on the utilization of the full 100 MHz bandwidth downlink PRS for positioning, this paper presents large-scale system-level simulations to illustrate the proposed 5G DPE positioning performance. Simulation results show that 5G DPE consistently achieves satisfactory positioning performance, with an overall RMSE constrained to within 6 m at an unsatisfactory SNR level as low as 10 dB. Furthermore, a comparative simulation conducted by using four candidate BSs shows that 5G DPE outperforms the OTDoA method by 95.24\% in terms of positioning accuracy in NLoS-dominated environments with imperfect synchronization.

\section{Acknowledgment}
This study was supported by the National Natural Science Foundation of China (NSFC) under Grant 62103346.

\end{document}